\documentclass[conference]{IEEEtran}
\IEEEoverridecommandlockouts
\usepackage{cite}
\usepackage{amsmath,amssymb,amsfonts}
\usepackage{graphicx}
\usepackage{textcomp}
\usepackage{xcolor}
\def\BibTeX{{\rm B\kern-.05em{\sc i\kern-.025em b}\kern-.08em
    T\kern-.1667em\lower.7ex\hbox{E}\kern-.125emX}}

\usepackage{booktabs}

\usepackage{array}
\usepackage{fix-cm}
\usepackage{times}
\usepackage{dirtytalk}
\usepackage{eurosym}
\usepackage[hyphens]{url}
\usepackage{hyperref}  
\usepackage{nomencl,etoolbox,ragged2e,siunitx}

\begin{document}
\title{Enhancing Gappy Speech Audio Signals with Generative Adversarial Networks\thanks{This work was partly supported by Science Foundation Ireland (SFI) under Grant Number SFI/12/RC/ 2289\_P2.}}
\author{\IEEEauthorblockN{Deniss Strods}
\IEEEauthorblockA{\textit{School of Computing} \\
\textit{Dublin City University}\\
Glasnevin, Dublin 9, Ireland \\
deniss.strods2@mail.dcu.ie}
\and
\IEEEauthorblockN{Alan F. Smeaton}
\IEEEauthorblockA{\textit{Insight Centre for Data Analytics} \\
\textit{Dublin City University}\\
Glasnevin, Dublin 9, Ireland  \\
alan.smeaton@dcu.ie}
}

\IEEEoverridecommandlockouts
\IEEEpubid{\makebox[\columnwidth]{
979-8-3503-4057-0/23/\$31.00~\copyright2023 IEEE\hfill} \hspace{\columnsep}\makebox[\columnwidth]{ }}

\maketitle              

\IEEEpubidadjcol

\begin{abstract}
Gaps, dropouts and short clips of corrupted audio are a common problem and particularly annoying when they occur in speech. This paper uses machine learning  to regenerate gaps of up to 320ms in an audio speech signal.  Audio regeneration is translated  into  image regeneration  by transforming audio  into a Mel-spectrogram and  using image in-painting   to regenerate the gaps. The full Mel-spectrogram is then transferred back to audio  using the Parallel-WaveGAN vocoder and integrated  into the audio stream.
Using a sample of 1300 spoken audio clips of between 1 and 10 seconds taken from the publicly-available LJSpeech dataset 
our results show  regeneration of audio gaps in close to real time using GANs
with a GPU equipped system.
As expected, the smaller the gap in the audio, the better  the quality of the filled gaps. On a gap of 240ms the 
average mean opinion score (MOS)  for the best performing models was 3.737, on a scale of 1 (worst) to 5 (best) which is sufficient for a human to perceive as close  to  uninterrupted human speech.
\end{abstract}
\section{Introduction}

Spoken audio can suffer from dropouts, gaps and short clips of corrupted data when transmitted over networks, including cellular networks.
This paper examines how generative adversarial networks (GANs), a form of machine learning  can enhance the quality of spoken audio by filling such gaps in real time. While there are classical machine learning approaches to enhance the quality of speech audio  based on Principal Component Analysis or  others that can clean an audio signal,  there is  no good approach  for real-time gap-filling.
Our approach is to transfer  audio regeneration  into  image in-painting  by converting gappy audio into to Mel-spectrograms, similar to work presented in \cite{zhou2019vision}. 
We examine data transmission packet loss conditions that produce gaps in audio varying from 40ms to 320ms, simulating a sequence of network packet losses of up to 8 packets. 

The next section  reviews  relevant research  covering GAN applications and variant architectures, and speech enhancement in noisy domains. Following that we present our experimental setup and then  our  results  followed by conclusions.

\section{Related Work}

\subsection{Speech Enhancement in  Noisy Audio}
Speech enhancement is an improvement task to the perceptual and aesthetic aspects of a speech signal which has been   degraded by noise. This task is performed in mobile communications, hearing aids and robust speech recognition \cite{phan2020improving, kim2021multi, pascual2017segan}. Even if the minimum required quality to understand what a person is saying is met, speech enhancement is still desirable as it can reduce listener fatigue.  The aesthetic enjoyment of listening to speech can be taken away due to  low fidelity of the speech in audio.  Simply increasing the fidelity of the speech signal may also boost the performance of speech-to-text algorithms \cite{loizou2007speech}.

Noise in a speech signal may come from a noisy communication channel or  the speech signal  may originate in a noisy location. In cases of voice-over-IP transmission, network packet loss is an issue that causes gaps in  transmission and reduces the perceptual and aesthetic features of a speech signal. Today, there is still no good solution available to the issue of regenerating gaps in audio signals in real-time communications. 

There are several approaches to  speech {\em enhancement}  including using principal component analysis, statistical model-based algorithms, spectral subtraction  and Wiener filtering. 
Recently speech enhancement has been addressed using  GANs \cite{phan2020improving,kim2021multi,pascual2017segan}, though in those works they train on either 462,880 utterances or 224,000 sentences, much greater than what is done here. GANs that work with audio and/or speech enhancement  typically use Mel-spectrograms, an image representation of an audio signal as shown later in Figure~\ref{fig:training_pairs}. A Mel-spectrogram  captures how  humans perceive sound better on lower frequencies compared to higher, and the spectrogram is a visualisation of the frequency composition of a signal over time. Features of this  can then be adjusted in order to improve the aesthetic or quality of the regenerated speech audio.

The existence of generative deep learning architectures, such as GANs,   allows us to address the  problem of {\em gappy} speech. A GAN's capability to generate from any complex data distribution suggests that a GAN may be trained to regenerate missing audio in real-time. The  data required to train such a model in a real setting may be collected from a speaker's previous speech and a model trained to regenerate gappy audio signals for that speaker. As part of a protocol among speakers, speech  models could be exchanged that would  be used  to enhance an incoming speech signal by resolving gaps in communication due to packet loss.

\subsection{Generative Adversarial Networks (GANs)}

Generative Adversarial Networks (GANs) are an approach to generative modelling using deep learning first introduced  by Goodfellow {\em et al.} 
in 2014 \cite{goodfellow2014generative}. 
Generative models 
allow learning a distribution of data without the need for extensively annotated training data. Based on  training data, GANs  allow generating new data similar to its training set. 

GAN architecture is based on  game theory, where backpropagation signals are derived through a competitive process. Two neural networks, a Generator $(G)$ and Discriminator $(D)$ compete with each other. $G$ learns to model distributions of data by trying to deceive  $D$ to recognise the generated samples as real \cite{goodfellow2020generative}. What is particularly useful is that  GAN models can be trained to mimic any distribution of data, so there are  many practical applications yet to be discovered \cite{alqahtani2021applications}.

The application of GANs was initially limited to image enhancement tasks like producing high-quality images, until about 2017 when the first GAN capable of facial image {\em generation} was created. GANs attracted attention and now we see  GANs used  where  synthetic data generation is required including natural language Processing \cite{croce2020gan}, computer vision \cite{caelles2019fast} and  audio generation \cite{donahue2018synthesizing}.

For some applications, it is difficult to train a GAN using the original GAN architecture as some generators do not learn the distribution of training data well enough and so the Deep Convolutional GAN (DCGAN) was proposed in 2015 \cite{radford2015unsupervised}. 
In this architecture, instead of the fully connected multi-layer perceptron NNs, CNNs were used. The authors in \cite{radford2015unsupervised} identified a sub-set of CNNs that were suitable for use in the GAN framework. To stabilise the training process, the generator used the ReLU activation function across the layers, except in the final layer, where the Tanh function was used. 
Some specific constraints on the model  identified during the development of the DCGAN laid the foundation of many further GAN architectures  based on DCGAN. These include the Conditional Generative Adversarial Network (cGAN)  \cite{mirza2014conditional} which can include labelling, WaveGAN which is used for audio synthesis  \cite{donahue2018synthesizing} and Parallel WaveGAN \cite{yamamoto2020parallel}  also used in audio and which uses auxiliary input features in the form of the Mel-spectrogram. 

For the speech {\em enhancement}, the Speech Enhancement GAN (SEGAN) architecture was introduced in \cite{pascual2017segan}. The generator network is used to perform enhancement of the signal, its input being the noisy signal and latent representation, and its output is the enhanced signal. The generator is structured in the same way as the auto-encoder. Encoding involves a number of strided convolutional layers followed by parametric rectified linear units (PReLUs), where the result of every N steps of the filter is a convolution. The discriminator plays the role of  expert classifier and conveys if the distribution is real or fake and the generator adjusts the weights towards the realistic distribution. 

As GANs are well developed in the areas of image-to-image translation and image in-painting  
\cite{jam2021r, wang2018image, zhou2019vision, isola2017image}, 
which are similar to gap regeneration tasks, we  propose to transform an audio signal into a Mel-spectrogram and use an image in-painter to fill the image gap. 
Mel-spectrograms can  then be in-painted and transferred back to audio via a neural vocoder, such as Parallel-WaveGAN \cite{yamamoto2020parallel}. 
We propose to train a model to regenerate the gap in the fixed position at the end of the Mel-spectrogram, which would make this problem simpler to tackle for a GAN as it would always know where to in-paint the image.

\section{Experimental Setup}
\label{sec:expts}

\subsection{Dataset}
The public domain LJSpeech data-set   \cite{ljspeech17} is used  which consists of 13,100 single-speaker short audio clips where the speaker  reads passages from 7 non-fiction books in  English. The entire duration of the clips, which range in length from 1 to 10 seconds,  is approximately 24 hours, and the dataset consists of 13,821 distinct words. For our experiments  a random subset of 1,300 clips was used with 1,000  used for training and 300 for testing. Our reason for using a  sub-set is to more closely represent a real world use case where less training data is available for a given voice requiring gap-filling in audio telephony and video conferencing. As mentioned earlier, related work such as  \cite{phan2020improving,kim2021multi,pascual2017segan} trains on either 462,880 utterances or 224,000 sentences.

\subsection{Data Pre-Processing}
\label{subsec:preproc}

Before the original 22kHz audio clips were converted via short-time Fourier transform (STFT) into Mel-spectrograms \cite{durak2003short}, the signal was trimmed at the start and end to remove silence. Thereafter STFT was performed on the audio  with a frame length of 1024 points (corresponding to 46ms) and a hop size of 256 points (11ms). STFT peaks were elevated by square function, to highlight voice pitch and then transformed to Mel scale using 80-channel characteristics. An additional parameter of Mel filterbank as frequency was set to include audio in the range from 80Hz to 7.6kHz. 

Mel-spectrograms were  scaled to have an approximately constant energy per channel, followed by log10 dynamic range compression. They were normalised by subtracting the global mean (µ) of the dataset  then dividing by the standard deviation $(\sigma)$. As a final step, values were normalised to the range  [-1,1]. In order to perform  normalisation,  statistics from the overall dataset were collected. The length of the clips was  standardised to 256 frames in the time domain (corresponding to 2.8s).

To mimic faulty communications typical of packet-based IP,  Mel-spectrograms with audio gaps  from 40ms  to  320ms were created at the end of the Mel-spectrogram as the real-time nature of audio communication requires regeneration to be applied as quickly as possible.  Thus a trailing audio signal following the gap  is not  available. The 40ms to 320ms gaps allow  mimicing of packet loss of up to 8 packets in a row, with the assumption that audio compression  captures 40ms of audio in one packet. Gaps longer than 320ms  introduce a risk of generating words that were not said because typical word rate for fast speech is up to 160 words per minute \cite{yuan2006towards} (375ms each) so this  sets the upper target for our  gap-filling. 

\begin{figure}[htb]
	\centering
		\includegraphics[width=0.49\textwidth]{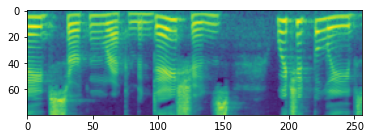}
		\includegraphics[width=0.49\textwidth]{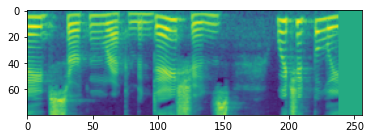}
	\caption{Training pairs used as input to our model, ground truth (top) and masked Mel with gap in green (bottom). 
	\label{fig:training_pairs}}
\end{figure}

The complete dataset is  formed from Mel-spectrogram pairs of source (Mel-spec with a gap) and target (ground truth) images. An example of a training pair is shown in Figure~\ref{fig:training_pairs}. The input to the model is the masked image and the model  tries to generate a complete image similar to the ground truth.

\subsection{Model and Loss Function}

The starting point for our in-painting  was Pix2Pix GAN \cite{isola2017image}. This was previously  used on multiple image-to-image transition tasks and  also used in similar work \cite{zhou2019vision}. In that related work the authors 
studied the creation of a joint feature space based on synchronised audio and video where the video consisted of spectrograms from the audio. That work focused on in-painting of the spectrogram to re-generate noisy or corrupted audio though their experiments were on music audio rather than on speech, which is our focus here.

\begin{figure}[htb]
	\centering
		\includegraphics[width=0.46\textwidth]{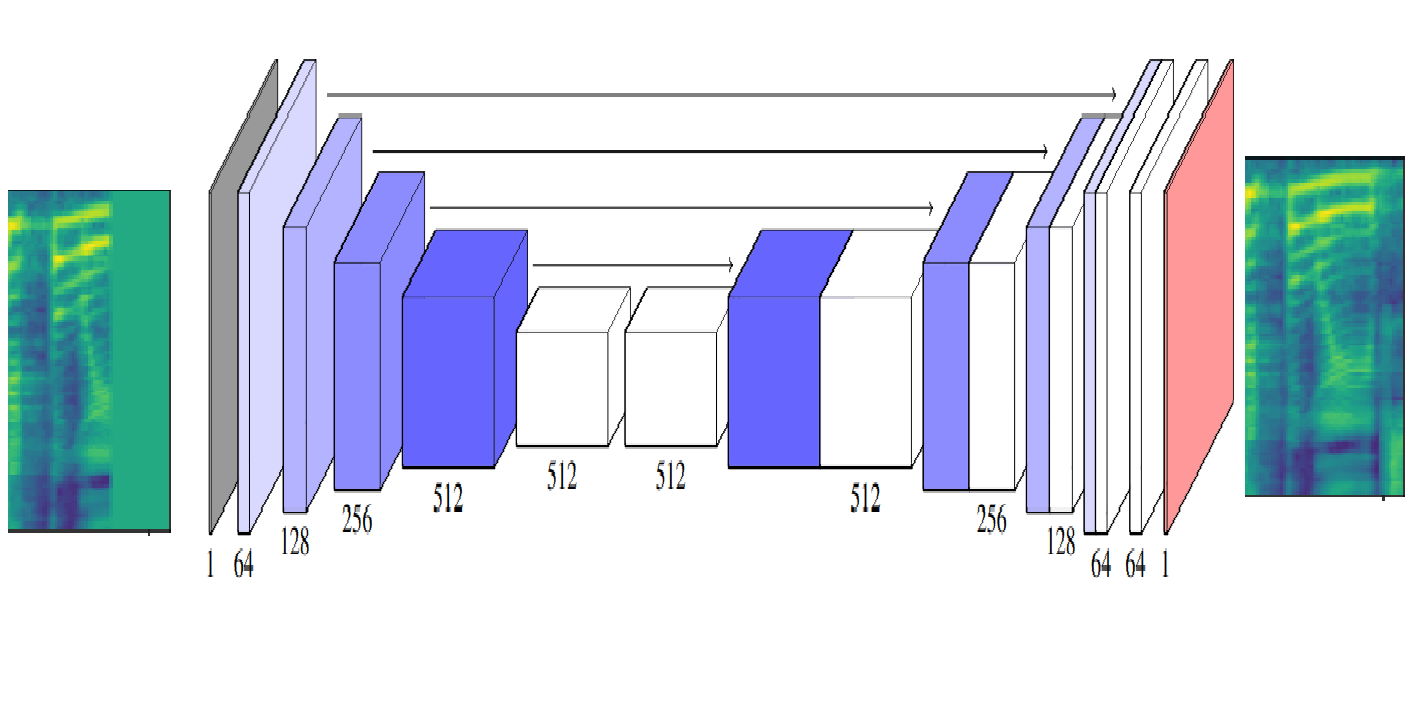}
		\hspace{0.05\textwidth}
		\includegraphics[width=0.46\textwidth]{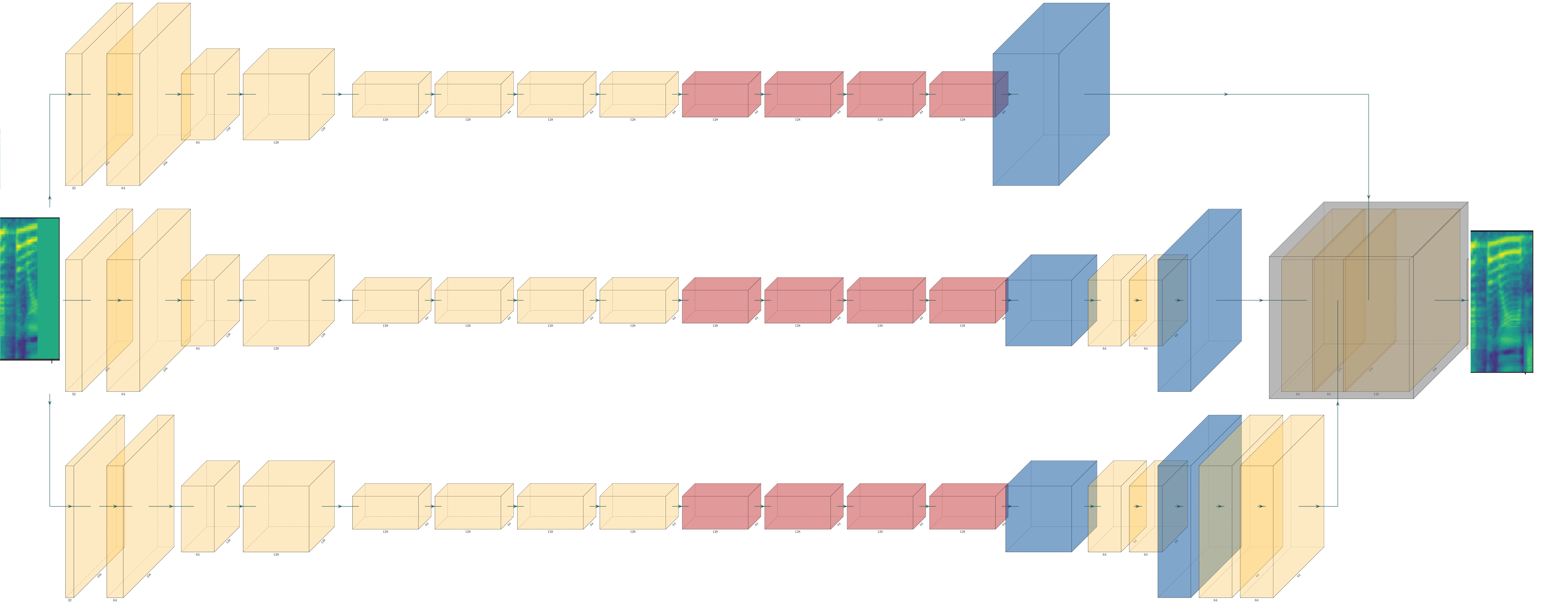}
	\caption{U-Net network (top) with skip connections denoted with the same colour blocks, GMCCN network (bottom).
	\label{fig:unet}\label{fig:GMCCN}}
\end{figure}

To form a baseline for this work,  a standard U-Net-based 5-layer generator presented in Figure~\ref{fig:unet} was used with L1 pixel-wise loss and input dimensions of 256x256. As part of the Pix2Pix architecture, the Patch-GAN discriminator was used for  adversarial loss, which creates scalar adversarial loss and Mean squared error (MSE) comparison of small patches of an image that form a grid and produce scores from 0 to 1, where each piece is classified as real or fake.

Alterations to the standard U-Net architecture Were performed. 
Stride configuration in the CNN layers was adjusted to allow   different size input dimensions of 125x128 and 256x80 pixels respectively to closely match our dataset profile. We also  used different variants of the loss functions by introducing  more advanced loss criteria   as proposed in more recent image in-painting work \cite{jam2021r, wang2018image}. We replaced the L1 pixel-wise loss with VGG19 feature match loss using VGG19 CNN's feature extraction layers to compare generated and ground truth images and updated gradients in the network based on comparative MSE error. VGG19 was pre-trained on Imagenet 
and the VGG19 feature match loss was  added to the in-painted segment.

The Generative Multi-column Convolutional Neural Network (GMCCN) \cite{wang2018image} was used in the same setting as the U-Net-based Generator. As shown in Figure~\ref{fig:GMCCN},  GMCCN  uses 3 networks in parallel and their output  is  concatenated at the final layer. We used the Patch-GAN discriminator for adversarial loss and VGG19 feature match loss for the GMCCN Generator. Modifications to the GMCCN were performed by introducing batch normalisation layers, as the network was susceptible to the exploding gradient problem as was discovered during our experiments.

The final stage of our pipeline was the Parallel-WaveGAN vocoder to convert Mel-spectrograms into waveforms. Typically Parallel-WaveGAN vocoder is used in the Tactron 2 Text to speech (TTS) pipeline \cite{yamamoto2020parallel}  where text is converted to Mel-spectrograms and vocoder generates audio and that may be used for any Mel-spectrogram to audio conversion. 
The
method used to train the Parallel WaveNet does not use any distillation process thus making the resulting model small and the overall processing fast.
A pre-trained Parallel-WaveGAN model pre-trained on the LJSpeech data was used here to avoid a costly training process. 

Our implementation was based on TensorFlow and trained on an NVIDIA GTX 1660 Super GPU. Networks were trained using the Adam optimiser 
with a learning rate set to  $1e-4$ and batch size  set to 1. All  data pre-processing, conversion to Mel-spectrograms and dataset matrix multiplications were computed via the TensorFlow API. Model performance  during   was recorded under Tensorboards.
The implementation of the Parallel-WaveGAN vocoder was based on  PyTorch, and weights were fetched from the public git repository at \url{https://github.com/kan-bayashi/ParallelWaveGAN}

Initial experimental models were trained for 40 epochs on the subset of 1,300 exemplars, with a fixed learning rate of $1e-4$ and beta of 0.5 set in the Adam optimiser. The default gap size was set to 240ms corresponding to 6 network packets. The gap was not variative in order to objectively assess different model performances in the same setting though later we present  experiments  with variative gap sizes using the best performing model.

\subsection{Evaluation Metrics}
We approach  evaluation  from the image aspect of the Mel-spectrograms and  from the audio aspect of the reconstructed WAV audio. Three  evaluation metrics are used. 
As a first measure, we  compute the mean squared error (MSE) of the pixels of the reconstructed image vs. the target image. As a second  metric, we  measure the MSE of the VGG19 CNN feature extraction layers of the Mel-spectrograms  and compare ground truth and generated data 
structures. We favour the VGG19 feature MSE metric over the L1 loss metric  as it is more descriptive visually, except in Table~\ref{tab:model_outputs}, which presents results in full. 
Because an image comparison metric does not clearly indicate how close to realistically sounding audio the generated in-painting actually generates, a third metric measures the quality of  generated audio  using the Perceptual Evaluation of Speech Quality (PESQ) \cite{de2008quality}
which is calculated for each test model. PESQ is a widely used standard for automated assessment of  speech  in telecommunication systems. It takes 2 audio samples as input and produces a Mean Opinion Score (MOS) from 1 (worst) to 5 (best).

\section{Experimental Results}
\label{sec:results}

The first  results reported are related to data normalisation. Our first test runs on the U-net architecture indicate that without  normalisation of the Mel-spectrograms, the model  fails to learn valid patterns and fails to produce  meaningful results. A set of normalisation techniques were applied that were described  in  Section~\ref{subsec:preproc}.  Evaluation results for the models are summarised in Table~\ref{tab:ed_data}.  

\begin{table*}[htb]
\centering
\caption{Summary of model output metrics}
\label{tab:ed_data}
        \begin{tabular}{rccc}
        \toprule 
        Metric & ~~~VGG19 Feature Loss~~~ &  ~~~~~~L1 Loss~~~~~~ & ~~~~~~MOS~~~~~~ \\
                \midrule 
        U-Net Baseline 256x128 & 6.056 & 0.415 & 2.348 \\
        U-Net 256x80 & 7.716 & 0.503 & 2.138 \\
        U-Net 128x128 & 9.962 & 0.636 & 2.029 \\
        U-Net  with VGG19 loss & 2.896 & 0.249 & 3.657 \\
        U-Net with VGG19 loss + chunk loss & 2.721 & 0.238 & 3.737 \\
        \midrule
        GMCCN & 3.402 & 0.255 & 2.465 \\
        \bottomrule
        \end{tabular}
\end{table*}

The baseline approach of in-painting with  normalised data shows  the algorithm is capable of learning the structure of the Mel-spectrogram and in-painting missing pieces with an MOS score of 2.348. However, its performance does not give the required result for a real life application. We  tried to match Mel-spectrogram dimensions to be closer to 256x80 pixels by changing the U-Net stride to 1 in the encoder-decoder connecting layers, however a rapid drop in map shrinking in the earlier layers caused a performance drop with MOS falling to 2.138. we identified that with minimal structural alteration the input size of 256x128  gave in-painting performance in line with the original 256x256. Thus all subsequent U-Net models had an input size of 256x128.

Following   recent in-painting approaches such as \cite{jam2021r, wang2018image}, we identified that newer approaches use more sophisticated loss functions and that  loss function alterations may boost  performance. Enhancements to our loss function, specifically VGG19 feature match loss for the whole image in addition to L1 loss were then applied. The in-painted image became closer to real data distribution and our VGG19 feature match error  decreased substantially from 6.056 to 2.896 and MOS increased from 2.348 up to 3.657. We also implemented an idea from \cite{jam2021r} where additional loss of the in-painted area was applied to the overall error. Therefore, the in-painted area loss was added to concentrate the attention of the algorithm more specifically on the in-painted area. That  decreased  VGG19 loss  further to 2.721  and increased MOS  to 3.737.

To understand whether the length of the Mel-spectrogram plays a role in predicting the masked segment,  input size was reduced from 256px (2.8s) to 125px (1.4s) by cutting Mel-spectrograms in half thus reducing the complexity of the problem as well as  computational cost. Training and testing found that reduction in data input significantly reduced the performance of the model. The VGG19 feature match score degraded from 6.056 (the baseline) to 9.962 indicating that the baseline algorithm  used information from the whole of the Mel-spectrogram. Experiments were performed around increasing the dimensionality of the data, but as that would have added additional computational cost, it was out of scope.

In addition to the U-Net generator, we conducted experiments with the GMCCN CNN architecture. 
The performance of GMCCN after our standard 40 training epochs was disappointing, with a VGG19 feature match loss of 3.402 and significant drop in MOS to 2.465. In addition, GMCCN has increased computational cost as the architecture includes 3 networks running in parallel as shown earlier in Figure~\ref{fig:GMCCN}.

\begin{figure}[htb]
	\centering
	\centerline{\includegraphics[width=0.4\textwidth]{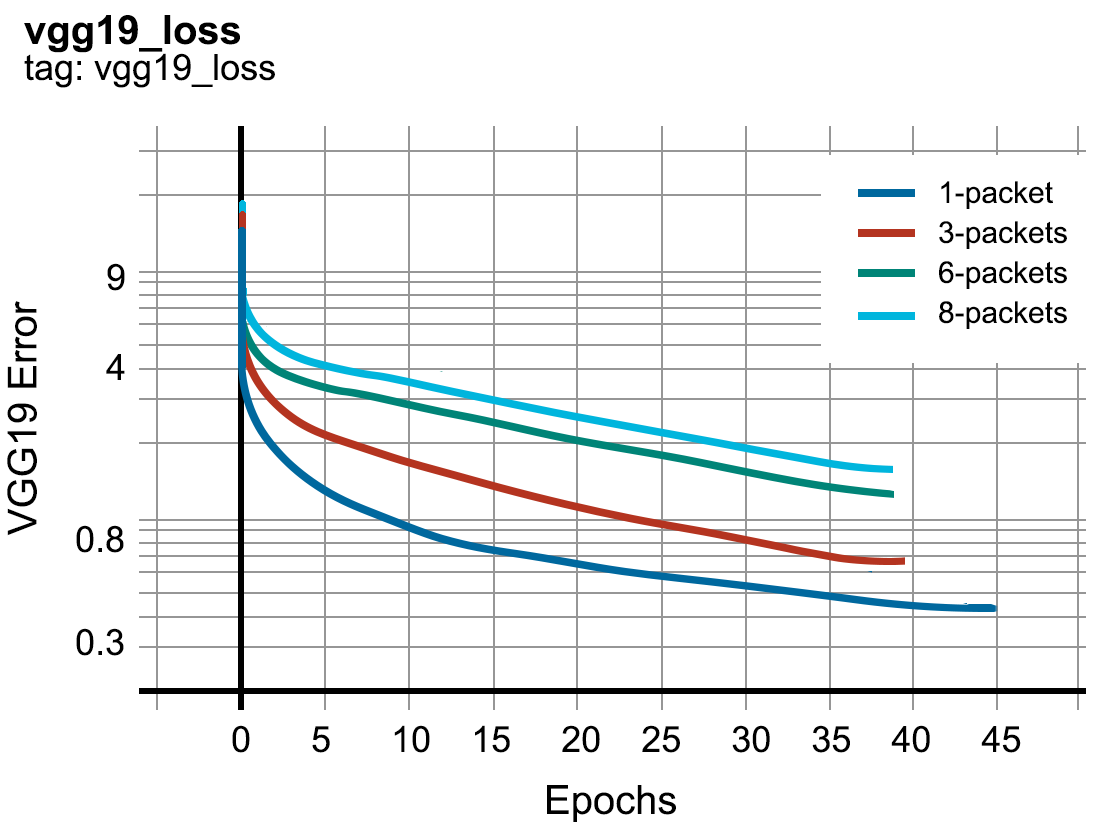}}
	\caption{Training loss with different gap size settings}\label{fig:multi_gap_raining_loss}
\end{figure}

To investigate the significance of the masked gap size, we conducted experiments based on the assumption that the algorithm would need to regenerate gaps from 40ms up to 320ms. Thus   models were  trained for different gap sizes. Results showed that reducing the gap size required less training time to achieve good performance as seen in Figure~\ref{fig:multi_gap_raining_loss}. An interesting finding was that when  Mel-spectrograms were generated on the 320ms gap model and others on the 160ms gap model,  the  error on the 320ms gap model  Mel-spectrogram was the same if we had taken the first 160ms of the sample. This tells us that  models perform the same if we for the same gap window.  Our subsequent experiments were carried out on segments with gaps of 320ms. Results also showed that performance degrades linearly and the model regenerates Mel-spectrograms with good confidence at the start, 
however the further into the time domain, the less accuracy results  as shown in Table~\ref{tab:different_packet_loss}.

\begin{table*}[htb]
\centering
\caption{Evaluation results for different mask sizes (in multiples of 40ms)}
\label{tab:different_packet_loss}
\begin{tabular}{rcccccccc}
\toprule
& 1&2&3&4&5&6&7&8\\
\midrule
MOS & ~~~~~~~4.514~~~ & 4.321~~~ & 4.185~~~ & 4.074~~~ & 3.938~~~ & 3.832~~~ & 3.520~~~ & 3.214 \\
VGG19 Loss & ~~~~0.501 & 0.989~~~ & 1.338~~~ & 1.832~~~ & 2.314~~~ & 2.762~~~ & 3.138~~~ & 3.738 \\
        \bottomrule
        \end{tabular}
\end{table*}

We also experimented with training models on variative gap sizes. In the data processing pipeline, random gap selection was performed in the range  40ms  to 320ms. After training the model for the default 40 epochs, the models that performed significantly worse than the fixed gap models were identified and the performance dropped by 70\% compared to fixed-size models. We  trained the model on a full data set of 13,000 samples (increasing the step amount from $40 \times 1000$  to $40 \times 13,000$). The model   still did not perform as well as those with fixed gap sizes, the VGG19 feature loss was 6.785, which is significantly higher than the 2.721 produced by the fixed gap model, even though  trained on a substantially larger dataset.

A series of tests of the inference speed on both the Parallel-WaveGAN and U-Net based generator Were  performed to identify if the model is usable in  real-time. The U-Net based model generates an in-painted Mel-spectrogram in approximately 50ms on a GPU,  in line with results presented in \cite{gadosey2020sd}. Parallel-WaveGAN converts a Mel-spectrogram to audio in 5ms on a GPU  in line with results presented in \cite{yamamoto2020parallel}. 

\begin{figure}[htb]
	\centering
		\includegraphics[width=0.36\textwidth]{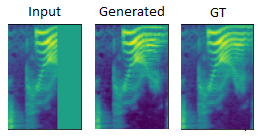}
				\hspace{0.05\textwidth}
		\includegraphics[width=0.36\textwidth]{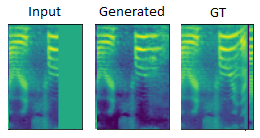}
	\caption{Example of the best (left) and worst (right) performing Mel-spectrograms as determined by VGG19 feature MSE. The black vertical line in the GT for the worst performing is a gap in the LJSpeech dataset sample audio} 
	\label{fig:best_worst}
\end{figure}

Finally, we  examined the worst performing in-painted Mel-spectrograms and best performing Mel-spectrogrms, identified by VGG19 feature loss and MOS. A summary of the results is shown in Table~\ref{tab:best_worst} and Figure~\ref{fig:best_worst} shows some representative examples.  A sample model output comparison may be seen in Table~\ref{tab:model_outputs}, along with  ground truth and the  Mel-spectrogram used as its input.

\begin{table}[htb]
\centering
\caption{Best and worst performing Mel-spectrogram results}
\label{tab:best_worst}
        \begin{tabular}{rcc}
        \toprule 
         & ~~~MOS~~~ & ~~VGG19 Loss\\
        \midrule 
        Best Performing by MOS  & 4.792 & 0.250 \\
        Best Performing by VGG19 / MSE  & 4.592 & 0.238 \\
        Worst Performing by MOS  & 2.735 & 3.871 \\
        Worst Performing by VGG19 / MSE  & 2.935 & 5.303 \\
        \bottomrule
        \end{tabular}
\end{table}

\begin{table*}[htb]
\centering
\caption{Comparing model outputs, Mel-spectrograms (size adjusted to fit)}
\label{tab:model_outputs}
        \begin{tabular}{cccc}
        \toprule 
         & Input & Ground Truth & \\
         \midrule
         & \includegraphics[width=70pt,height=70pt]{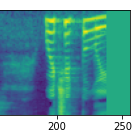} &         \includegraphics[width=70pt,height=70pt]{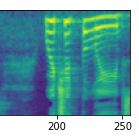} & \\
         \midrule
         U-Net Baseline & With VGG19 loss & With VGG19 loss & GMCNN \\
         &&+ chunk loss &\\
        \midrule 
        \includegraphics[width=70pt,height=70pt]{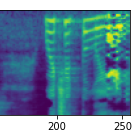} & 
        \includegraphics[width=70pt,height=70pt]{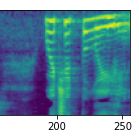} & 
        \includegraphics[width=70pt,height=70pt]{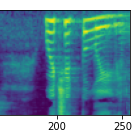} & 
        \includegraphics[width=70pt,height=70pt]{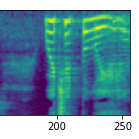} \\
        \bottomrule
        \end{tabular}
\end{table*}

\section{Conclusions} 

This paper presented a technique that improves a speech signal  degraded by the introduction of variable length gaps which arise frequently in in real-time audio telephony and  video conferencing.  Missing audio gaps are generated using GANs  as currently there is no good solution available for this task making comparison with prior work difficult. 
Our key findings are that U-Net-based GANs with a loss function based on VGG19 feature match \cite{wang2018image} for Mel-spectrograms from the audio are capable of in-painting gaps in those Mel-spectrograms in near real-time. 
After transforming an in-painted Mel-spectrogram back to audio via the Parallel-WaveGan vocoder \cite{yamamoto2020parallel} and following the use of an enhanced U-net generator with a more advanced loss function similar to one  in \cite{jam2021r, wang2018image}, we  generated audio fragments that are structurally  similar to the real distribution with a MOS from 3.214 for gaps of 320ms up to a MOS of 4.514 for gaps of 40ms.
The total time taken to regenerate a gap  is approximately 105ms on a GPU,  an acceptable performance for real-time communications.

For larger  regenerated gaps 
our model is  capable of almost exactly regenerating the missing area in the Mel-spectrogram. 
The model  uses information from all of the Mel-spectrogram, as  reducing the size of the input Mel-spectrogram  leads to a large drop in performance.
We found that fixed gap size models are capable of learning distributions from smaller data-sets as the complexity of the problem is reduced and  the most efficient way to address variative gap sizes is to train a model capable of filling large gaps and use it  for all gap sizes. The performance of such an approach is similar to that of models   trained on  smaller gap sizes.

We  conclude that it is possible to use our  in-painter-Vocoder pipeline to regenerate audio gaps in real-time on systems equipped with a GPU and that the result can be perceived by humans as good quality. 
Further work should identify if there are reduced sized models  similar to SD-UNET \cite{gadosey2020sd} that could perform well enough on CPU-only systems.

%
%
%
\bibliographystyle{splncs04}
\bibliography{issc23}

\begin{thebibliography}{10}
\providecommand{\url}[1]{\texttt{#1}}
\providecommand{\urlprefix}{URL }
\providecommand{\doi}[1]{https://doi.org/#1}

\bibitem{alqahtani2021applications}
Alqahtani, H., Kavakli-Thorne, M., Kumar, G.: Applications of generative
  adversarial networks ({GAN}s): An updated review. Archives of Computational
  Methods in Engineering  \textbf{28}(2),  525--552 (2021)

\bibitem{caelles2019fast}
Caelles, S., Pumarola, A., Moreno-Noguer, F., Sanfeliu, A., Van~Gool, L.: Fast
  video object segmentation with spatio-temporal {GANs}. arXiv preprint
  arXiv:1903.12161  (2019)

\bibitem{croce2020gan}
Croce, D., Castellucci, G., Basili, R.: Gan-bert: Generative adversarial
  learning for robust text classification with a bunch of labeled examples. In:
  58th Annual Meeting of the ACL. pp. 2114--2119 (2020)

\bibitem{donahue2018synthesizing}
Donahue, C., McAuley, J., Puckette, M.: {Synthesizing Audio with GANs}.
  \url{https://openreview.net/forum?id=r1RwYIJPM} (2018), last Accessed
  16-Feb-2023

\bibitem{durak2003short}
Durak, L., Arikan, O.: Short-time {F}ourier transform: two fundamental
  properties and an optimal implementation. IEEE Transactions on Signal
  Processing  \textbf{51}(5),  1231--1242 (2003)

\bibitem{gadosey2020sd}
Gadosey, P.K., Li, Y., Agyekum, E.A., Zhang, T., Liu, Z., Yamak, P.T., Essaf,
  F.: {SD-UNET}: Stripping down {U}-net for segmentation of biomedical images
  on platforms with low computational budgets. Diagnostics  \textbf{10}(2),
  ~110 (2020)

\bibitem{goodfellow2014generative}
Goodfellow, I., Pouget-Abadie, J., Mirza, M., Xu, B., Warde-Farley, D., Ozair,
  S., Courville, A., Bengio, Y.: Generative adversarial nets. Advances in
  neural information processing systems  \textbf{27} (2014)

\bibitem{goodfellow2020generative}
Goodfellow, I., Pouget-Abadie, J., Mirza, M., Xu, B., Warde-Farley, D., Ozair,
  S., Courville, A., Bengio, Y.: Generative adversarial networks. Comm. ACM
  \textbf{63}(11),  139--144 (2020)

\bibitem{isola2017image}
Isola, P., Zhu, J.Y., Zhou, T., Efros, A.A.: Image-to-image translation with
  conditional adversarial networks. In: IEEE Conference on Computer Vision and
  Pattern Recognition. pp. 1125--1134 (2017)

\bibitem{ljspeech17}
Ito, K., Johnson, L.: The {LJ} speech dataset.
  \url{https://keithito.com/LJ-Speech-Dataset/} (2017), last Accessed: 11-Feb,
  2023

\bibitem{jam2021r}
Jam, J., Kendrick, C., Drouard, V., Walker, K., Hsu, G.S., Yap, M.H.: {R-mnet}:
  A perceptual adversarial network for image inpainting. In: IEEE/CVF Winter
  Conference on Applications of Computer Vision. pp. 2714--2723 (2021)

\bibitem{kim2021multi}
Kim, H.Y., Yoon, J.W., Cheon, S.J., Kang, W.H., Kim, N.S.: A multi-resolution
  approach to {GAN}-based speech enhancement. Applied Sciences  \textbf{11}(2),
  ~721 (2021)

\bibitem{de2008quality}
de~Lima, A.A., Freeland, F.P., de~Jesus, R.A., Bispo, B.C., Biscainho, L.W.,
  Netto, S.L., Said, A., Kalker, A., Schafer, R., Lee, B., et~al.: On the
  quality assessment of sound signals. In: IEEE International Symposium on
  Circuits and Systems. pp. 416--419 (2008)

\bibitem{loizou2007speech}
Loizou, P.C.: Speech enhancement: theory and practice. CRC Press (2007)

\bibitem{mirza2014conditional}
Mirza, M., Osindero, S.: Conditional generative adversarial nets. arXiv
  preprint arXiv:1411.1784  (2014)

\bibitem{pascual2017segan}
Pascual, S., Bonafonte, A., Serra, J.: {SEGAN: Speech enhancement generative
  adversarial network}. arXiv preprint arXiv:1703.09452  (2017)

\bibitem{phan2020improving}
Phan, H., McLoughlin, I.V., Pham, L., Ch{\'e}n, O.Y., Koch, P., De~Vos, M.,
  Mertins, A.: Improving {GANs} for speech enhancement. IEEE Signal Proc.
  Letters  \textbf{27},  1700--1704 (2020)

\bibitem{radford2015unsupervised}
Radford, A., Metz, L., Chintala, S.: Unsupervised representation learning with
  deep convolutional generative adversarial networks. arXiv preprint
  arXiv:1511.06434  (2015)

\bibitem{wang2018image}
Wang, Y., Tao, X., Qi, X., Shen, X., Jia, J.: Image inpainting via generative
  multi-column convolutional neural networks. In: Advances in Neural
  Information Processing Systems. pp. 331--340 (2018)

\bibitem{yamamoto2020parallel}
Yamamoto, R., Song, E., Kim, J.M.: Parallel {WaveGAN}: A fast waveform
  generation model based on generative adversarial networks with
  multi-resolution spectrogram. In: International Conference on Acoustics,
  Speech and Signal Processing (ICASSP). pp. 6199--6203 (2020)

\bibitem{yuan2006towards}
Yuan, J., Liberman, M., Cieri, C.: Towards an integrated understanding of
  speaking rate in conversation. In: Ninth International Conference on Spoken
  Language Processing (2006)

\bibitem{zhou2019vision}
Zhou, H., Liu, Z., Xu, X., Luo, P., Wang, X.: Vision-infused deep audio
  inpainting. In: Proceedings of the IEEE/CVF Conference on Computer Vision.
  pp. 283--292 (2019)

\end{thebibliography}

\end{document}